\documentclass[fleqn,usenatbib]{mnras}

\usepackage{newtxtext,newtxmath}
\usepackage[T1]{fontenc}

\DeclareRobustCommand{\VAN}[3]{#2}
\let\VANthebibliography\thebibliography
\def\thebibliography{\DeclareRobustCommand{\VAN}[3]{##3}\VANthebibliography}

\usepackage{graphicx}	% Including figure files
\usepackage{amsmath}	% Advanced maths commands
\title[White dwarfs made of heavy elements]{Ultra low-mass and small-radius white dwarfs made of heavy elements}

\author[C.-J. Xia et al.]{
Cheng-Jun~Xia,$^{1}$\thanks{E-mail: cjxia@yzu.edu.cn}
Yong-Feng~Huang,$^{2, 3}$\thanks{E-mail: hyf@nju.edu.cn}
Hong-Bo~Li,$^{4,5}$\thanks{E-mail: lihb2020@stu.pku.edu.cn}
Lijing~Shao$^{5,6}$\thanks{E-mail: lshao@pku.edu.cn}
and Ren-Xin~Xu$^{4,5}$\thanks{E-mail: r.x.xu@pku.edu.cn}
\\$^{1}${Center for Gravitation and Cosmology, College of Physical Science and Technology, Yangzhou University, Yangzhou 225009, China}
\\$^{2}${School of Astronomy and Space Science, Nanjing University, Nanjing 210023, China}
\\$^{3}${Key Laboratory of Modern Astronomy and Astrophysics (Nanjing University), Ministry of Education, Nanjing 210023,  China}
\\$^{4}${School of Physics, Peking University, Beijing 100871, China}
\\$^{5}${Kavli Institute for Astronomy and Astrophysics, Peking University, Beijing 100871, China}
\\$^{6}${National Astronomical Observatories, Chinese Academy of Sciences, Beijing 100012, China}
}

\date{Accepted XXX. Received YYY; in original form ZZZ}

\pubyear{2015}

\begin{document}
\label{firstpage}
\pagerange{\pageref{firstpage}--\pageref{lastpage}}
\maketitle

% Abstract of the paper
\begin{abstract}
Seven ultra low-mass and small-radius white dwarfs (LSPM J0815+1633, LP 240-30, BD+20 5125B, LP 462-12, WD J1257+5428, 2MASS J13453297+4200437, and SDSS J085557.46+053524.5) have been recently identified with masses ranging from $\sim$0.02 $M_\odot$ to $\sim$0.08 $M_\odot$ and radii from $\sim$ 4270 km to 10670 km. The mass-radius measurements of these white dwarfs pose challenges to traditional white dwarf models assuming they are mostly made of nuclei lighter than $^{56}$Fe. In this work we consider the possibility that those white dwarfs are made of heavier elements. Due to the small charge-to-mass ratios in heavy elements, the electron number density in white dwarf matter is effectively reduced, which reduces the pressure with additional contributions of lattice energy and electron polarization corrections. This consequently leads to white dwarfs with much smaller masses and radii, which coincide with the seven ultra low-mass and small-radius white dwarfs. The corresponding equation of state and matter contents of dense stellar matter with and without reaching the cold-catalyzed ground state are presented, which are obtained using the latest Atomic Mass Evaluation (AME 2020). Further observations are necessary to unveil the actual matter contents in those white dwarfs via, e.g., spectroscopy, asteroseismology, and discoveries of other ultra low-mass and small-radius  white dwarfs.
\end{abstract}

% Select between one and six entries from the list of approved keywords.
% Don't make up new ones.

%%%%%%%%%%%%%%%%%%%%%%%%%%%%%%%%%%%%%%%%%%%%%%%%%%

%%%%%%%%%%%%%%%%% BODY OF PAPER %%%%%%%%%%%%%%%%%%

\section{\label{sec:intro}Introduction}
White dwarfs represent the final destiny of the vast majority of stars, which may reach temperature $\sim$100 eV and density $\sim$10$^6$ g/cm$^3$ in their centers~\citep{Saumon2022_PR988-1}. The matter contents of typical white dwarfs are $^{12}$C and $^{16}$O, covered by a thin envelope of $^{4}$He (and $^{1}$H if not burned entirely). If the progenitor star approaches to the 10$M_\odot$ limit, white dwarfs are thought to be made of $^{16}$O and $^{20}$Ne~\citep{Siess2007_AA476-893}, while He white dwarfs are also possible if the progenitors are very low-mass and in a binary system~\citep{Iben1993_PASP105-1373, Marsh1995_MNRAS275-828}. As white dwarfs slowly cool down, a crystallized core will be formed, releasing latent heat that delays the cooling process~\citep{Tremblay2019_Nature565-202}. Most white dwarfs are expected to go through at least one pulsation phase during their evolution, displaying periodic variations in their brightness that arise from global oscillations~\citep{Fontaine2008_PASP120-1043, Corsico2019_AAR27-7}. Additionally, various oscillation modes can be excited during the late inspiral or merger of white dwarf binaries, which may emit gravitational waves that are detectable for future space-borne gravitational wave detectors~\citep{Tang2023_MNRAS521-926}.

\begin{table*}
\caption{\label{table:sWD} Masses, radii, and surface temperatures of seven ultra low-mass and small-radius white dwarfs~\citep{Rebassa-Mansergas2016_MNRAS458-3808, Blouin2019_ApJ878-63, Kurban2022_PLB-137204}.}
\begin{tabular}{l|ccc} \hline \hline
MWDD ID                  &         $M$       &        $R$              &   $T_\mathrm{eff}$      \\
                         &     $M_\odot$     &        km               &      K                  \\ \hline
LSPM J0815+1633          & 0.082 $\pm$ 0.031 & 13563.23 $\pm$ 1024.76  &   4655 $\pm$ 35        \\
LP 240-30                & 0.081 $\pm$ 0.016 & 13542.5  $\pm$ 626.6    &   4680 $\pm$ 25         \\
BD+20 5125B              & 0.08  $\pm$ 0.038 & 13046.72 $\pm$ 1124.18  &   4395 $\pm$ 90         \\
LP 462-12                & 0.054 $\pm$ 0.024 & 11999.23 $\pm$ 1552.78  &   4800 $\pm$ 20         \\
WD J1257+5428            & 0.032 $\pm$ 0.03  & 12403.13 $\pm$ 3561.25  &   7485 $\pm$ 85         \\
2MASS J13453297+4200437  & 0.031 $\pm$ 0.04  & 9186.23  $\pm$ 3405.51  &   4270 $\pm$ 75         \\
SDSS J085557.46+053524.5 & 0.02  $\pm$ 0.245 & 14688.29 $\pm$ 767.07   &  10670 $\pm$ 1677       \\
\hline
\end{tabular}
\end{table*}

Combined with the measurements on the distance and surface temperature $T_\mathrm{eff}$ of a white dwarf~\citep{Blouin2019_ApJ878-63}, its radius $R$ can be inferred according to the observed flux. The surface gravity of a white dwarf can also be measured according to the gravitational redshift of the spectrum lines from atmosphere~\citep{Fontaine2001_PASP113-409, GentileFusillo2018_MNRAS482-4570, Chandra2020_ApJ899-146}, which can be used to fix the mass with additional information on its radius. Throughout the available data on the masses and radii of white dwarfs in the Montreal White Dwarf Database~\citep[MWDD; ][]{Dufour2017}, as indicated in Table~\ref{table:sWD}, seven ultra low-mass and small-radius white dwarfs have been identified with the masses ranging from $\sim$0.02~$M_\odot$ to $\sim$0.08 $M_\odot$ and radii from $\sim$ 4270 km to 10670 km~\citep{Kurban2022_PLB-137204}. This poses challenges to traditional white dwarf models, which are considered to be made of light elements such as $^{12}$C, $^{16}$O, $^{4}$He, and $^{20}$Ne. Consequently, traditional white dwarf models predict much larger radii than those indicated in Table~\ref{table:sWD}.

To understand the physical origin of such low-mass and small-radius white dwarfs, possible candidates made of various exotic matter were considered. For example, \cite{Kurban2022_PLB-137204} proposed that they are in fact strange dwarfs comprised of a strange quark matter core and a thick normal matter crust~\citep{Glendenning1995_PRL74-3519}. It was suggested that the intermittent fractional collapses of the crust induced by refilling of materials accreted from its low-mass companion lead to repeating fast radio bursts~\citep{Geng2022_Inov2-100152}. Additionally, there are also possibilities that those white dwarfs may be strangelet dwarfs~\citep{Alford2012_JPG39-065201} or $ud$QM dwarfs~\citep{Wang2021_Galaxies9-70, Xia2022_PRD106-034016} comprised of strangelets or $ud$QM nuggets emersed in a sea of electrons.

In this work we consider the possibility that those low-mass and small-radius white dwarfs are made of heavy elements such as $^{56}$Fe, $^{62}$Ni, $^{108}$Pd, and $^{208}$Pb, which reduces significantly the mass and radius of a white dwarf in comparison with those made of light elements. In fact, there exist extensive observations of (DZ) white dwarfs contaminated by heavy elements, which was identified by the characteristic spectral lines~\citep{Farihi2016_NAR71-9, Zuckerman2018, Coutu2019_ApJ885-74}. It is thus reasonable to consider the possible existence of low-mass white dwarfs made entirely of heavy elements, where the shell of light elements is either stripped by its companion object in a binary system or by nuclear explosion.

The paper is organized as follows. In Sec.~\ref{sec:the} we present the theoretical framework for obtaining the properties of white dwarf matter under various constraints on mass numbers of nuclei, including either light or heavy elements. The obtained equation of state (EOS) and matter contents that minimize the energy density of cold white dwarf matter are presented in Sec.~\ref{sec:results}, while the corresponding white dwarf structures are investigated and compared with the low-mass and small-radius white dwarfs. We draw our conclusion in Sec.~\ref{sec:con}

\section{Theoretical framework}\label{sec:the} 
For cold white dwarf matter comprised of crystallized nuclei emersed in a sea of electrons, the energy density can be divided into three parts~\citep{Chamel2020_PRC101-032801}, i.e.,
\begin{equation}
  E = \frac{M_N n_e}{Z} + \left(1+\frac{\alpha}{2 \pi}\right) E_e + K_M \alpha \left(\frac{4\pi n_e^4 Z^2}{3}\right)^{1/3}\sigma(Z), \label{eq:Et}
\end{equation}
where $n_e$ is the average electron number density and $\alpha=1/137.03599911$ the fine structure constant. Here the first term represents the energy density of nuclei with $M_N(Z, A)$ being the mass of a nucleus with $Z$ protons and $A$ nucleons, which is determined by
\begin{equation}
M_N(Z, A) = M_A(Z, A) - Zm_e + B_e(Z)
\end{equation}
with $M_A(Z, A)$ being the measured atomic mass~\citep[AME 2020; ][]{Huang2021_CPC45-30002, Wang2021_CPC45-030003},  $m_e = 510998.95$ eV the electron mass, and $B_e(Z) = \big( 14.4381 Z^{2.39}+1.55468\times 10^{-6} Z^{5.35} \big) $ eV the electron binding energy~\citep{Lunney2003_RMP75-1021}. The baryon number density is then $n_\mathrm{b}=n_e/f_Z$ with the charge-to-mass ratio $f_Z=Z/A$. The second term in Eq.~(\ref{eq:Et}) corresponds to the energy density of electrons including exchange contributions, where $E_e$ is the energy density of free electron gas, i.e.,
\begin{equation}
E_e = \frac {{m_e}^4}{8\pi^{2}} \left[x_e(2x_e^2+1)\sqrt{x_e^2+1}-\mathrm{arcsh}(x_e) \right]
\end{equation}
with $x_e\equiv {(3\pi^2 n_e)^{1/3}}/{m_e}$. The third term in Eq.~(\ref{eq:Et}) represents the lattice energy density including electron polarization corrections. For a body-centered cubic lattice, the Madelung constant $K_M = -0.895929255682$~\citep{Baiko2001_PRE64-57402} and the function $\sigma(Z)$~\citep{Chamel2020_PRC101-032801, Potekhin2000_PRE62-8554} is given by
\begin{equation}
\sigma(Z) = 1 + \frac{ 12^{{4}/{3}}  Z^{{2}/{3}}\alpha}{35 \pi^{{1}/{3}}} \left(1-\frac{ 1.1866}{Z^{ 0.267}}+\frac{ 0.27}{Z}\right).
\end{equation}

For nuclei to stably exist inside white dwarfs, they should be stable against electron capture and $\beta$-decay  reactions, i.e.,
\begin{eqnarray}
  && {}^{A}_{Z}X + e^- \rightarrow  {}^{A}_{Z-1}X + \nu_e; \\
  && {}^{A}_{Z}X  \rightarrow  {}^{A}_{Z+1}X + e^- + \bar{\nu}_e.
\end{eqnarray}
This indicates the following stability condition, i.e.,
\begin{equation}
  E(Z\pm 1, A, n_\mathrm{b}) - E(Z, A, n_\mathrm{b}) >0 \label{eq:bstbl}
\end{equation}
with the energy density $E$ fixed by Eq.~(\ref{eq:Et}). Note that the electron number density changes into $n_e(Z\pm 1)/Z$ as we vary the charge number of a nucleus from $Z$ to $Z\pm 1$. A vast number of nuclei species fulfilling the stability condition~(\ref{eq:bstbl}) is obtained, which could all exist stably inside white dwarfs if there are no other decay channels. At fixed baryon number density $n_\mathrm{b}$, we search for the nucleus that minimizes the energy density of white dwarf matter under three different considerations, i.e.,
\begin{enumerate}
  \item Massive white dwarfs comprised of light elements ($A\leq 16$, 24, 28);
  \item Catalyzed ones with all possible nuclear species;
  \item Ultra low-mass and small-radius white dwarfs made of heavy elements ($A\geq 62$, 108, 208).
\end{enumerate}
Once the nucleus is fixed, the energy density can then be obtained with Eq.~(\ref{eq:Et}). Similar procedure is carried out in a vast density range with $n_\mathrm{b}\approx 10^{-13}\mbox{--}10^{-4}$ fm$^{-3}$. At larger densities, the nucleus becomes too neutron-rich so that neutrons start to drip out and form a neutron gas, which is beyond the scope of current study. According to basic thermodynamic relations, the pressure of white dwarf matter is then determined by
\begin{equation}
  P = \left(1+\frac{\alpha}{2 \pi}\right) P_e + K_M \alpha \left(\frac{4\pi n_e^4 Z^2}{3^4}\right)^{1/3}\sigma(Z), \label{eq:Pt}
\end{equation}
with $P_e=m_e\sqrt{x_e^2+1}n_e-E_e$.

At smaller densities with pressure $P\lesssim 0.4P_e \approx 10^{-17}$ MeV/fm$^3$, nuclei are not fully ionized as few electrons start to bound to them. In such cases, the EOS predicted by Eqs.~(\ref{eq:Et}) and (\ref{eq:Pt}) is no longer valid. We follow the treatment of~\cite{Baym1971_ApJ170-299} and employ the results obtained by~\cite{Feynman1949_PR75-1561}, where the pressure of white dwarf matter now becomes
\begin{equation}
  P = \frac{3^{2/3} \pi^{4/3} }{5 m_e} \left[f(Z, n_e) n_e\right]^{5/3}. \label{eq:Po}
\end{equation}
Note that a dampening factor $f(Z, n_e)$ is introduced, where Eq.~(\ref{eq:Po}) becomes the pressure of noninteracting electrons in the non-relativistic limit if we take $f=1$. The exact value of $f(Z, n_e)$ is fixed by interpolating the results presented in the Fig.~1 of~\citet{Feynman1949_PR75-1561}.

\section{Results and Discussion}\label{sec:results}

\begin{figure}
\includegraphics[width=\linewidth]{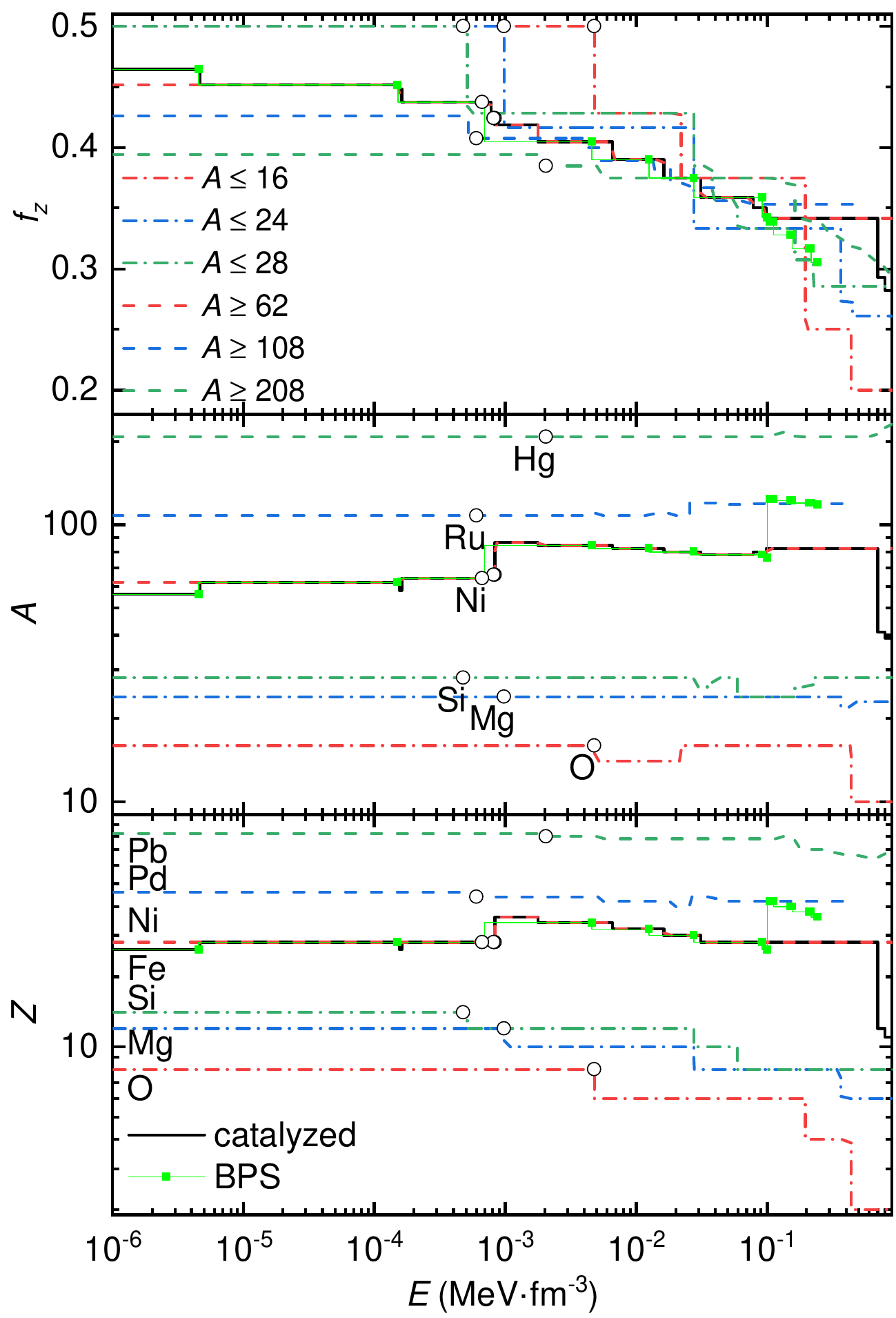}
\caption{\label{Fig:Nucleon} Proton number $Z$, mass number $A$, and charge-to-mass ratio $f_Z=Z/A$ of nuclei in white dwarf matter under different considerations, where the latest atomic mass evaluation~\citep[AME 2020; ][]{Huang2021_CPC45-30002, Wang2021_CPC45-030003} has been employed. For comparison, the nuclear species from BPS EOS is presented~\citep{Baym1971_ApJ170-299}. The open circles indicate the central densities of the most massive white dwarfs as in the sixth column of Table~\ref{table:WD}.}
\end{figure}

In Fig.~\ref{Fig:Nucleon} we present the proton number $Z$, mass number $A$, and charge-to-mass ratio $f_Z=Z/A$ of nuclei in white dwarf matter as functions of energy density, where various constraints on the mass numbers of nuclei are adopted with $A\leq 16$, 24, 28 or $A\geq 62$, 108, 208. Similar to the BPS model~\citep{Baym1971_ApJ170-299}, the catalyzed one indicted by the black solid curve is obtained by searching for all possible nuclear species without any restriction on $A$, which generally agrees with the BPS EOS at $E\lesssim 0.1$ MeV/fm$^{3}$. The latest data from the atomic mass evaluation (AME 2020) have been adopted in our calculations~\citep{Huang2021_CPC45-30002, Wang2021_CPC45-030003}. It is found that the nuclear species remains unchanged at $E\lesssim 3\times10^{-4}$ MeV/fm$^{3}$ except for the catalyzed one, covering most of the density range in white dwarfs. As indicated by the symbols in the bottom panel of Fig.~\ref{Fig:Nucleon} as well as in the second column of Table~\ref{table:WD}, at small densities the nuclei ${}^{16}$O, ${}^{24}$Mg, ${}^{28}$Si, ${}^{56}$Fe, ${}^{62}$Ni, ${}^{108}$Pd, and ${}^{208}$Pb minimize the energy density of white dwarf matter under various constraints on $A$. The corresponding charge-to-mass ratio $f_Z$ starts to decrease with $A$ at $A\geq 62$, which reduces the electron number density with $n_e=f_Z n_\mathrm{b}$. As will be shown later, this will consequently reduces the pressure and makes white dwarfs more compact. At larger densities, the nuclear species starts to change, which typically takes place at the center of the most massive white dwarfs (indicated by the open circles in Fig.~\ref{Fig:Nucleon}) except for the catalyzed one. If we further increase the density, the nuclear species continues to change which decreases the charge-to-mass ratio $f_Z$.

\begin{table*}
\caption{\label{table:WD} Summary of white dwarf properties obtained under different constraints on nuclear mass numbers. The radii ($R_{0.03}$) of 0.03$M_\odot$ white dwarfs and their matter contents ($^{A}$$X_{0.03}$) are indicated in the third and second columns. The masses ($M_\mathrm{max}$) and radii ($R_\mathrm{max}$) of the most massive white dwarfs are presented, along with the energy density $E_\mathrm{c}$, pressure $P_\mathrm{c}$, and nuclei species ($^{A}$$X_\mathrm{c}$) at the center of these white dwarfs.}
\begin{tabular}{c|cc|ccccc} \hline \hline
Criterion  &  $^{A}$$X_{0.03}$   & $R_{0.03}$ & $M_\mathrm{max}$& $R_\mathrm{max}$ & $E_\mathrm{c}$ & $P_\mathrm{c}$ &  $^{A}$$X_\mathrm{c}$     \\
           &            &     km     &   $M_\odot$     &        km        &  eV/fm$^3$  &  eV/fm$^3$  &              \\ \hline
$A\leq 16$ &  $^{16}$O  &   21300    &       1.38      &        1422      &     4790.2     &      4.283     &  $^{16}$O   \\
$A\leq 24$ &  $^{24}$Mg &   19767    &       1.33      &        2201      &      976.4     &      0.489     &  $^{24}$Mg   \\
$A\leq 28$ &  $^{28}$Si &   19123    &       1.30      &        2522      &      477.9     &      0.238     &  $^{28}$Si   \\ \hline
Catalyzed  &  $^{56}$Fe &   15137    &       1.00      &        2080      &      830.3     &      0.384     &  $^{66}$Ni   \\ \hline
$A\geq 62$ &  $^{62}$Ni &   14300    &       1.00      &        2064      &      811.4     &      0.383     &  $^{66}$Ni   \\
$A\geq108$ & $^{108}$Pd &   11760    &       0.89      &        2087      &      601.5     &      0.238     & $^{108}$Ru   \\
$A\geq208$ & $^{208}$Pb &    9130    &       0.75      &        1391      &     2040.7     &      1.098     & $^{208}$Hg   \\
\hline
\end{tabular}
\end{table*}

\begin{figure}
\includegraphics[width=\linewidth]{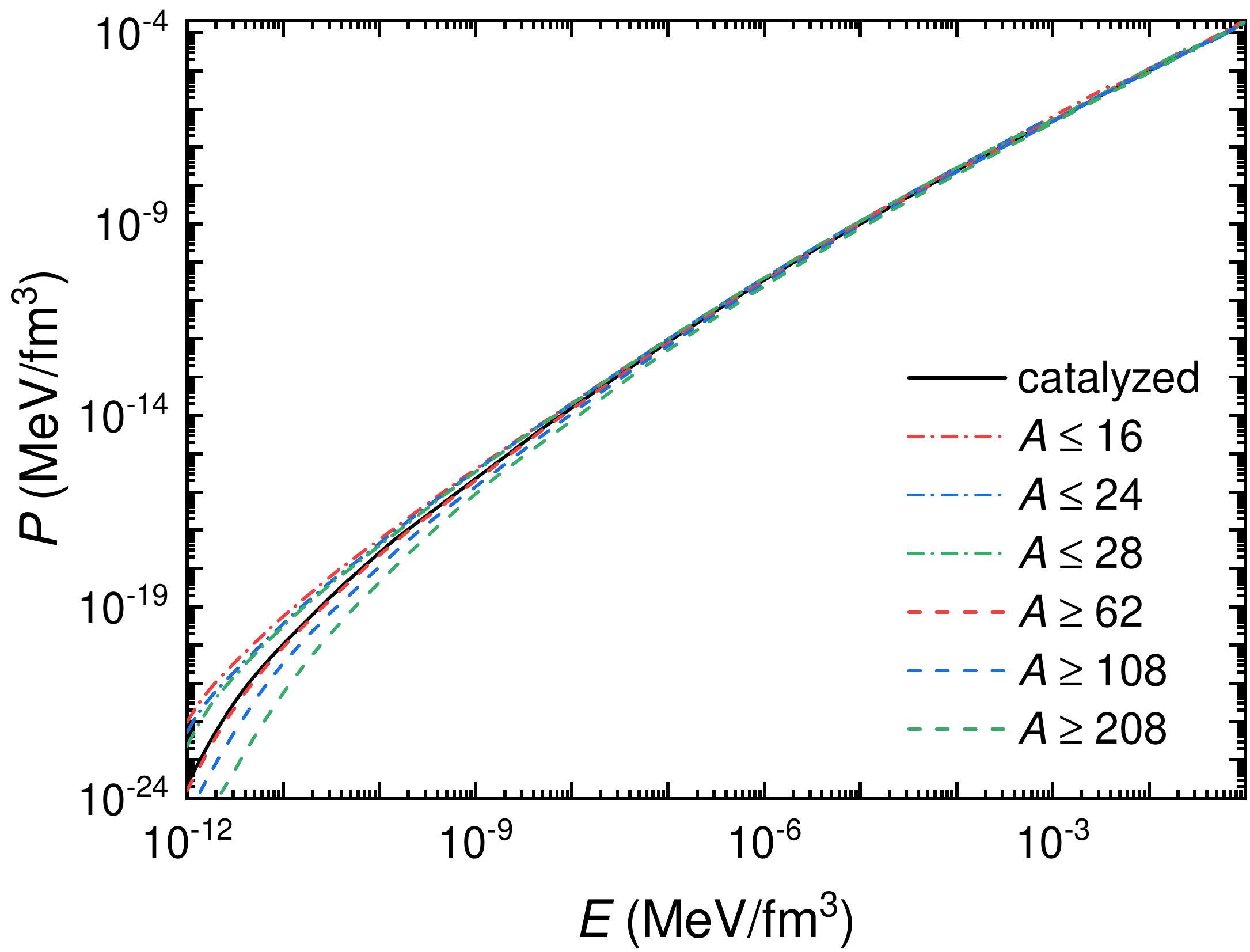}
\caption{\label{Fig:EOS} Pressure of white dwarf matter as functions of energy density, where the corresponding nuclear species is indicated in Fig.~\ref{Fig:Nucleon}.}
\end{figure}

Based on the nuclear species indicated in Fig.~\ref{Fig:Nucleon}, the pressure of white dwarf matter can then be fixed using Eqs.~(\ref{eq:Pt}) and (\ref{eq:Po}). The corresponding EOSs under various constraints are then presented in Fig.~\ref{Fig:EOS}. At $E \gtrsim 10^{-4}$ MeV/fm$^{3}$, the EOSs of white dwarf matter generally coincide with each other, while there are slight variations due to the sudden changes in nuclear species causing mild first-order phase transitions. At $E \lesssim 10^{-4}$ MeV/fm$^{3}$, the effects of chemical composition become evident, where the pressure is effectively reduced if white dwarf matter is made of heavy elements. This is attributed to the reduction of charge-to-mass ratio $f_Z$ at $A\geq 62$, which decreases the electron number density and consequently the pressure. The pressure reduction becomes more evident at smaller densities, where the variations in lattice energy density and electron polarization corrections become sizable even for cases with same $f_Z$.

\begin{figure}
\includegraphics[width=\linewidth]{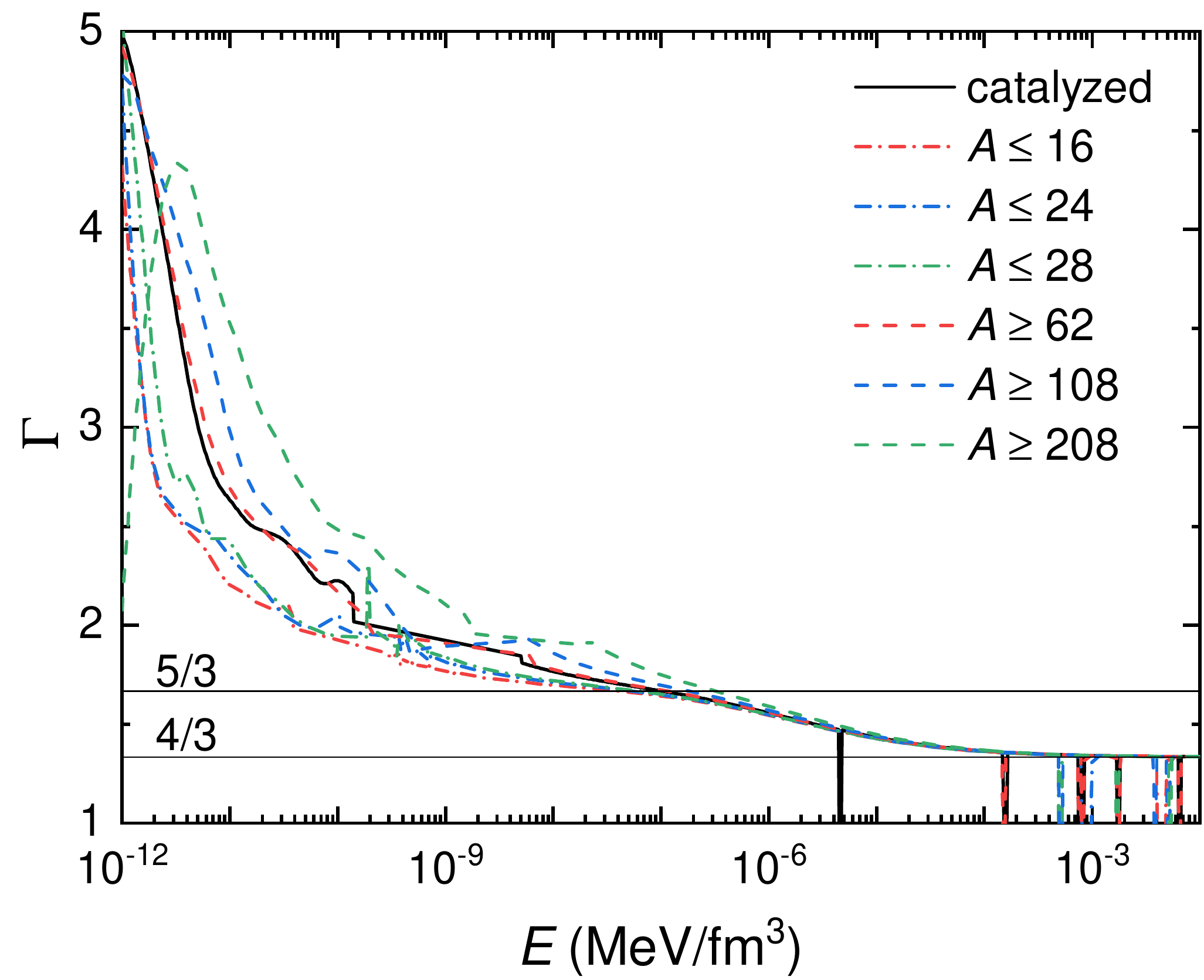}
\caption{\label{Fig:Gamma} Adiabatic index of white dwarf matter as functions of energy density.}
\end{figure}

To better illustrate the effects of heavy elements on the EOSs of white dwarf matter, in Fig.~\ref{Fig:Gamma} we present the corresponding adiabatic index as functions of energy density, which is determined by
\begin{equation}
  \Gamma = \frac{\mbox{d}\ln P}{\mbox{d}\ln n_\mathrm{b}}.
\end{equation}
For electron gas in the non-relativistic limit, one expects $\Gamma = 5/3$, which turns into 4/3 in the extreme relativistic limit with a softer EOS. This is indeed the case for white dwarf matter at $E \gtrsim 10^{-4}$ MeV/fm$^{3}$, where the EOSs in Fig.~\ref{Fig:EOS} generally coincide with each other with $\Gamma = 4/3$. Nevertheless, there are few exceptions during the first-order phase transitions with sudden changes in nuclear species, which reduces the adiabatic index to $\Gamma = 0$. At smaller densities, however, $\Gamma$ easily exceeds the non-relativistic limit $5/3$. In particular, as density decreases, $\Gamma$ quickly increases and becomes larger if white dwarf matter is made of heavy elements with larger $A$. This can be attributed to the additional contributions of lattice energy and electron polarization corrections, which are sizable at small densities.

\begin{figure}
\includegraphics[width=\linewidth]{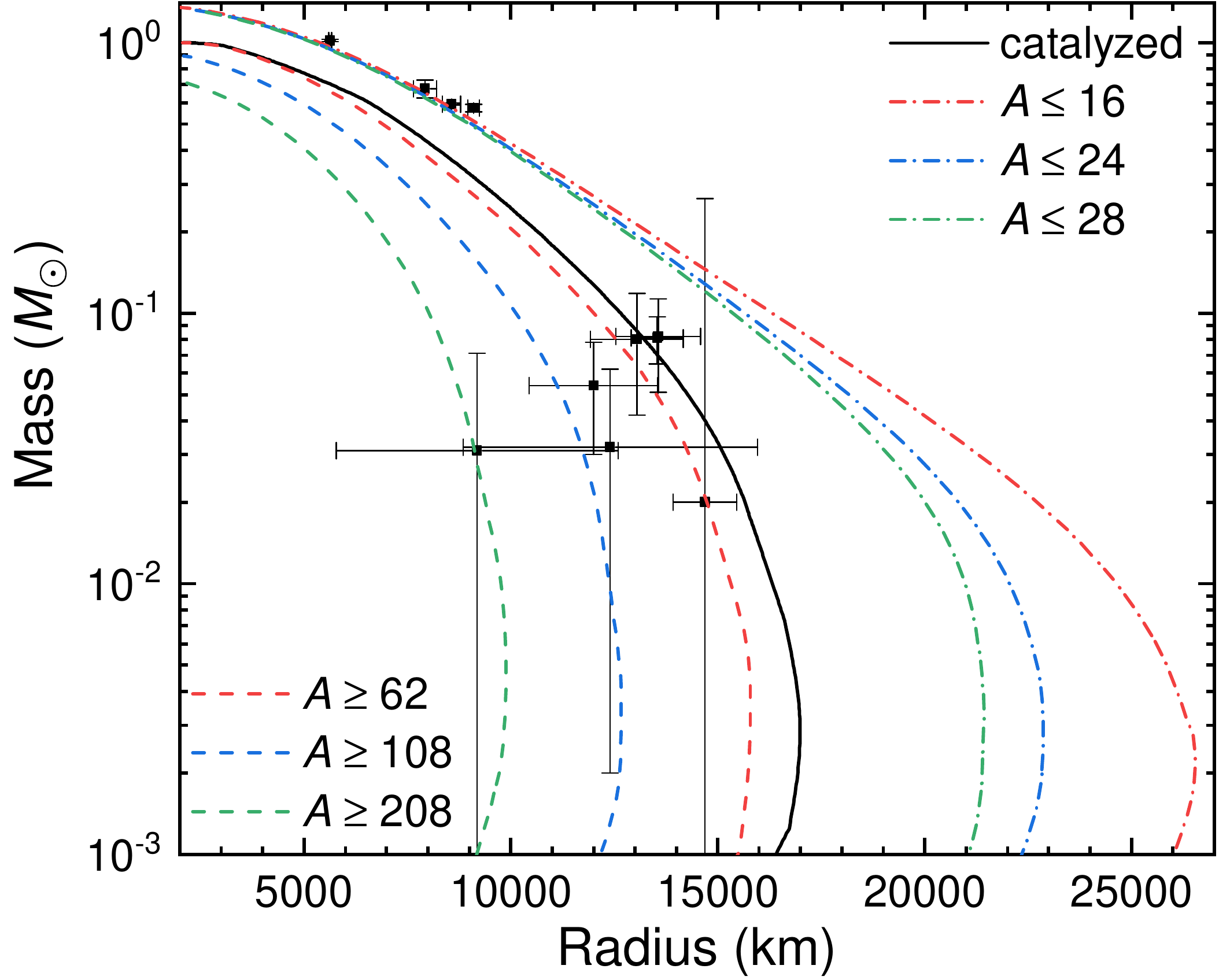}
\caption{\label{Fig:MR} Mass-radius relations of white dwarfs obtained with the EOSs presented in Fig.~\ref{Fig:EOS}. The dots with $M>0.5 M_\odot$ represent four typical white dwarfs Sirius B, Stein 2051 B, Procyon B, and 40 Eri B~\citep{Bond2017_ApJ848-16}, while the seven dots below correspond to the ultra low-mass and small-radius white dwarfs LSPM J0815+1633, LP 240-30, BD+20 5125B, LP 462-12, WD J1257+5428, 2MASS J13453297+4200437, and SDSS J085557.46+053524.5~\citep{Kurban2022_PLB-137204}.}
\end{figure}

Based on the EOSs presented in Fig.~\ref{Fig:EOS}, the structures of white dwarfs are fixed by solving the TOV equation
\begin{eqnarray}
&&\frac{\mbox{d}P}{\mbox{d}r} = -\frac{G m E}{r^2}   \frac{(1+P/E)(1+4\pi r^3 P/m)} {1-2G m/r},  \label{eq:TOV}\\
&&\frac{\mbox{d}m}{\mbox{d}r} = 4\pi E r^2, \label{eq:m_star}
\end{eqnarray}
where $G=6.707\times 10^{-45}\ \mathrm{MeV}^{-2}$ is the gravitational constant. In Fig.~\ref{Fig:MR} we present the mass-radius relations of white dwarfs under various constraints on their matter contents. Typical white dwarfs with $M>0.5 M_\odot$~\citep{Bond2017_ApJ848-16} and seven ultra low-mass and small-radius white dwarfs are indicated by the dots with the corresponding error bars~\citep{Kurban2022_PLB-137204}. It is evident that the typical white dwarfs are made of light elements, where the mass and radius could become slightly larger if lighter elements ($^{12}$C, $^{4}$He, $^{1}$H), magnetic field, and temperature effects are accounted for. If white dwarfs are made of heavier elements, as indicated in Fig.~\ref{Fig:MR}, the mass and radius decrease significantly, which coincide with the observed seven ultra low-mass and small-radius white dwarfs LSPM J0815+1633, LP 240-30, BD+20 5125B, LP 462-12, WD J1257+5428, 2MASS J13453297+4200437, and SDSS J085557.46+053524.5~\citep{Kurban2022_PLB-137204}. This is attributed to the reduction of pressure at small densities if white dwarfs are made of heavy elements, which become more compact than typical white dwarfs.

\begin{figure}
\includegraphics[width=\linewidth]{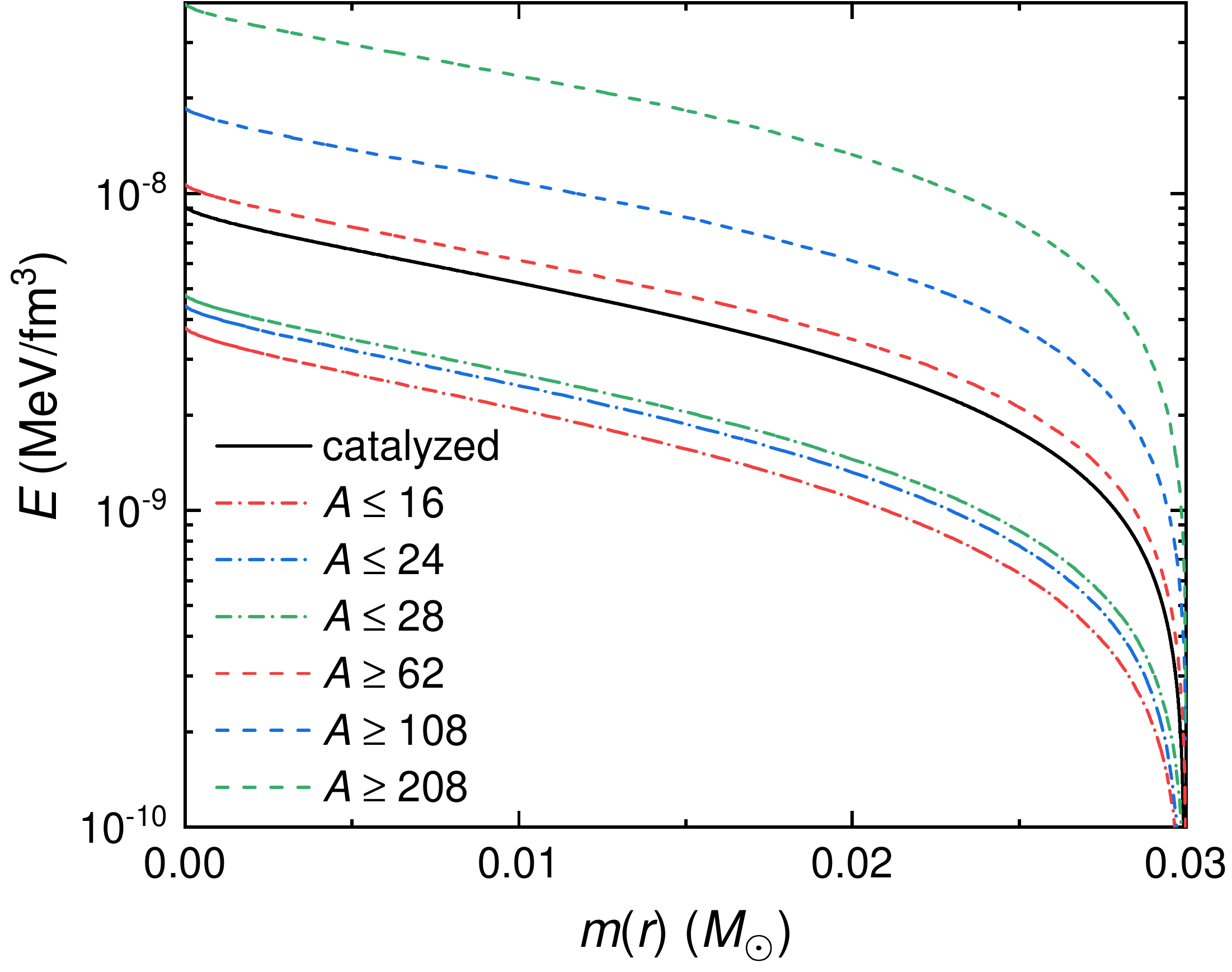}
\caption{\label{Fig:Int} Energy density profiles of $0.03 M_\odot$ white dwarfs as functions of the total mass $m(r)$ enclosed in a sphere of radius $r$ in Eq.~(\ref{eq:m_star}), where the corresponding nuclear species and radius are indicated in the second and third columns of Table~\ref{table:WD}.}
\end{figure}

To show this explicitly, in Fig.~\ref{Fig:Int} we present the internal energy density profiles of $0.03 M_\odot$ white dwarfs, where the horizontal axis corresponds to the total mass $m(r)$ enclosed in a sphere of radius $r$ in Eq.~(\ref{eq:m_star}). The corresponding nuclear species that makes up the white dwarf and its radius are indicated in the second and third columns  of Table~\ref{table:WD}. It is evident that the energy density of white dwarf matter increases with the nuclear mass number $A$, leading to more compact white dwarfs with smaller radii.

\section{Conclusion}\label{sec:con}
The recent observations of the seven ultra low-mass and small-radius white dwarfs LSPM J0815+1633, LP 240-30, BD+20 5125B, LP 462-12, WD J1257+5428, 2MASS J13453297+4200437, and SDSS J085557.46+053524.5 have posed challenges to traditional white dwarf models assuming that they are mostly made of nuclei lighter than $^{56}$Fe. In this work we consider the possibility that those white dwarfs are made of heavier elements, which effectively reduces the pressure and leads to white dwarfs with much smaller masses and radii. Further observations are necessary to unveil the actual matter contents in those white dwarfs via, e.g., their cooling processes~\citep{Tremblay2019_Nature565-202}, characteristic spectral lines emitted by the heavy elements~\citep{Farihi2016_NAR71-9, Zuckerman2018, Coutu2019_ApJ885-74}, pulsation that arises from their global oscillations~\citep{Fontaine2008_PASP120-1043, Corsico2019_AAR27-7}, gravitational waves excited during the late inspiral or merger of white dwarf binaries~\citep{Tang2023_MNRAS521-926}, and searching for other white dwarfs with similar low-mass and small-radius characteristics~\citep{Kurban2022_PLB-137204}.

\section*{ACKNOWLEDGMENTS}
This work was supported by National Natural Science Foundation of China (Grants No.~12275234, No.~12233002, and No.~12041306), National SKA Program of China (Grants No.~2020SKA0120300 and No.~2020SKA0120100).

%12275234      致密星物质相变及其微观结构的研究      夏铖君       扬州大学    55            2023.01-2026.12

%%%%%%%%%%%%%%%%%%%% REFERENCES %%%%%%%%%%%%%%%%%%

% The best way to enter references is to use BibTeX:

%\bibliographystyle{mnras}
%\bibliography{strange_quark} % if your bibtex file is called example.bib

% Alternatively you could enter them by hand, like this:
% This method is tedious and prone to error if you have lots of references
%\begin{thebibliography}{99}
%\bibitem[\protect\citeauthoryear{Author}{2012}]{Author2012}
%Author A.~N., 2013, Journal of Improbable Astronomy, 1, 1
%\bibitem[\protect\citeauthoryear{Others}{2013}]{Others2013}
%Others S., 2012, Journal of Interesting Stuff, 17, 198
%\end{thebibliography}

% Don't change these lines
\bsp	% typesetting comment
\label{lastpage}
\end{document}